\documentclass[lettersize,journal]{IEEEtran}
\usepackage{amsmath,amsfonts}
\usepackage{algorithmic}
\usepackage{algorithm}
\usepackage{array}
\usepackage[caption=false,font=normalsize,labelfont=sf,textfont=sf]{subfig}
\usepackage{textcomp}
\usepackage{stfloats}
\usepackage{url}
\usepackage{verbatim}
\usepackage{graphicx}
\usepackage{cite}
\usepackage{dsfont}
\usepackage{multirow}
\usepackage{tabularx}
\usepackage{booktabs}
\usepackage{pifont}

\newcommand{\cmark}{\ding{51}} 
\newcommand{\xmark}{\ding{55}} 

\begin{document}

\title{PRMB: Benchmarking Reward Models in Long-Horizon CBT-based Counseling Dialogue}

\author{Yougen Zhou, Qin Chen, Ningning Zhou, Jie Zhou, and Liang He
\thanks{Yougen Zhou is with the Shanghai Institute of Artificial Intelligence for Education, East China Normal University, Shanghai, China (e-mail: zyg@stu.ecnu.edu.cn).}
\thanks{Ningning Zhou is with the School of Psychology and Cognitive Science, East China Normal University, Shanghai, China (e-mail: nnzhou@psy.ecnu.edu.cn).}
\thanks{Qin Chen, Jie Zhou, and Liang He are with the School of Computer Science and Technology, East China Normal University, Shanghai, China (e-mail: qchen@cs.ecnu.edu.cn). Qin Chen is the corresponding author.}
}

\markboth{Journal of \LaTeX\ Class Files,~Vol.~14, No.~8, August~2021}%
{Shell \MakeLowercase{\textit{et al.}}: A Sample Article Using IEEEtran.cls for IEEE Journals}


\maketitle

\begin{abstract}
Large language models (LLMs) hold potential for mental healthcare applications, particularly in cognitive behavioral therapy (CBT)-based counseling, where reward models play a critical role in aligning LLMs with preferred therapeutic behaviors. However, existing reward model evaluations often fail to capture alignment effectiveness in long-horizon interventions due to limited coverage of process-oriented datasets and misalignment between evaluation targets and psychological alignment objectives. To address these limitations, we present PRMB, a comprehensive benchmark tailored for evaluating reward models in multi-session CBT counseling. PRMB spans 6 sessions and 21 diverse negative scenarios, incorporating both pairwise and Best-of-N preference evaluations. We demonstrate a positive correlation between our benchmark and downstream counseling dialogue performance. Based on our benchmark, we conduct extensive analysis on the state-of-the-art reward models, revealing their generalization defects that were not discovered by previous benchmarks and highlighting the potential of generative reward models. Furthermore, we delve into examining the effectiveness of inference-time strategy for the evaluation of reward models and analyzing the impact factors of generative reward models. This work advances intelligent informatics for personalized healthcare by establishing a framework for reward model assessment in mental health dialogues. Evaluation code and datasets are publicly available at \url{https://github.com/YouKenChaw/PRMB}. 
\end{abstract}


\section{Introduction}
\label{sec:introduction}
\IEEEPARstart{L}{arge} language models (LLMs) are increasingly being adopted in mental health-related scenarios, including emotional support conversation, cognitive restructuring, and psychotherapy dialogues \cite{qiu-etal-2024-smile, chen-etal-2023-soulchat, he-etal-2025-ecc, cr}. Among established therapeutic paradigms, Cognitive Behavioral Therapy (CBT) is one of the most structured and evidence-based approaches, making it a natural target for LLM-based simulated counseling systems \cite{lee-etal-2024-cactus}. Effective CBT-based counseling requires not only linguistic fluency and helpfulness, but also safety, therapeutic appropriateness, and consistency across sessions. However, evaluating whether an LLM-based simulated counselor satisfies these requirements remains an open challenge. Recent work has explored using LLM-based judge agents to evaluate responses under predefined criteria \cite{xie-etal-2025-psydt, zhang-etal-2024-cpsycoun}. These judge models are often further adapted as reward models (RMs) to guide reinforcement learning in training simulated counselor agents \cite{Zhang2025SentientAA, Wang2025RLVERRL}. Despite their growing use, the reliability of RMs in counseling-oriented settings remains insufficiently validated.
\begin{figure}[!t]
\centerline{\includegraphics[width=\columnwidth]{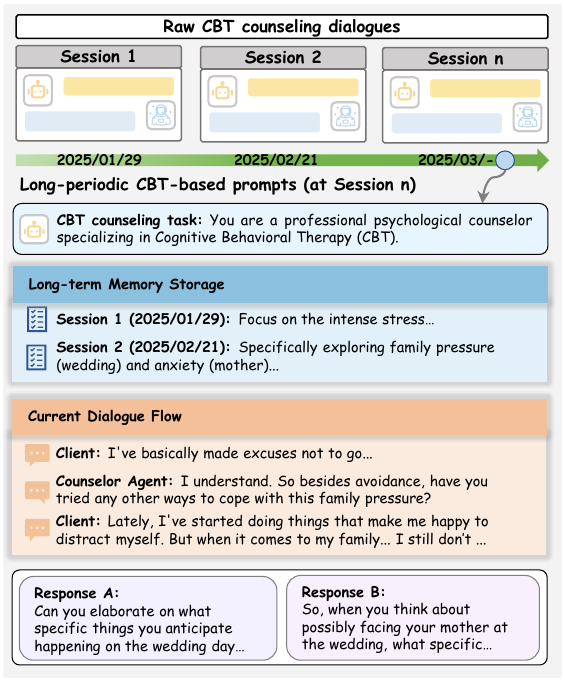}}
\caption{CBT-based counseling dialogues require session-level coherence, long-term consistency across multiple sessions, and adherence to structured therapeutic progression.}
\label{fig:intruction}
\end{figure}

Most existing RM benchmarks focus on short-context general dialogue settings, typically involving single-turn or brief multi-turn exchanges that emphasize isolated response quality \cite{lambert-etal-2025-rewardbench, Zhou2024RMBCB, Malik2025RewardBench2A}. In contrast, CBT-based counseling dialogues require session-level coherence, long-term consistency across multiple sessions, and adherence to therapeutic progression, as illustrated in Fig.~\ref{fig:intruction}. As a result, RM signals derived from short-horizon benchmarks may fail to capture critical process violations that only emerge over extended counseling trajectories, leaving a critical gap in validating RMs for long-horizon counseling tasks.

To address this gap, we introduce \textbf{PRMB}, a benchmark for evaluating RMs in long-horizon, multi-session CBT-based counseling. PRMB is constructed using publicly available CBT counseling cases and carefully crafted simulated client personas solely for research and evaluation. For each counseling case, we adopt a progressive summarization framework that incrementally integrates historical information with the current session context, enabling realistic long-context evaluation without exposing raw dialogue histories. Based on this setup, we curate over $12$k pairwise preference instances and Best-of-N samples generated by ten representative LLMs.

Using PRMB, we investigate three key dimensions. First, we examine whether existing RMs can generalize across diverse CBT counseling scenarios and long-term interaction contexts. Second, we examine the predictive power of our benchmark for downstream performance by performing Best-of-N sampling. Third, we conduct controlled analyses to examine how different inference-time strategies affect RM in long-horizon CBT counseling evaluation. Through extensive benchmarking and analysis, our results reveal substantial limitations of current RMs in CBT counseling, particularly in terms of consistency, robustness, and downstream alignment. We believe that PRMB provides a necessary foundation for systematically evaluating RMs in process-oriented conversational tasks. The main contributions of this paper are as follows:
\begin{itemize}
\item We introduce \textbf{PRMB}, a benchmark designed to evaluate reward models in long-horizon, multi-session CBT-based counseling dialogue.
\item We benchmark representative discriminative and LLM-as-judge reward models, revealing significant limitations in their consistency and robustness for counseling-oriented evaluation.
\item We conduct analyses of inference-time strategies across reward models to provide empirical insights on improving reward model reliability in long-horizon settings for counseling dialogue.
\end{itemize}

The remainder of this paper is organized as follows. Section \ref{sec:related_work} reviews related work on LLM-based simulated psychological dialogue, reward models, and existing benchmarks for reward model evaluation. Section \ref{sec:data_construction} details the construction process and statistics of our benchmark. Section \ref{sec:benchmarking} presents benchmarking results on state-of-the-art reward models. Section \ref{sec:downstream} assess the correlation between our benchmark
and downstream counseling dialogue. Section \ref{sec:discussion} discusses the effectiveness of inference-time strategy and analyzing the impact factors of reward models. Finally, Section \ref{sec:conclusion} concludes the paper and outlines directions for future work.

\section{Related Work}
\label{sec:related_work}
\subsection{Simulated Psychological Dialogue}
Large language models (LLMs) have advanced therapy-oriented dialogue by enabling more coherent and emotionally supportive interactions \cite{liu2023chatcounselor, chen2023soulchat, jo2023understanding, wei2024leveraging}. Subsequent studies incorporate counseling knowledge and structured therapeutic strategies to improve psychological support quality \cite{hsu2023helping, chen2023llm, na2024cbt}, emphasizing symptom identification and strategy-conditioned response generation. Recently, reinforcement learning (RL) has been introduced to align conversational agents with therapeutic objectives \cite{zhang2025echo, Wang2025RLVERRL}. Affect-aware reward signals are used to optimize empathetic behavior and perceived support quality. This shifts optimization from text similarity toward higher-level psychological outcomes, making reward models key components for behavioral alignment in therapeutic dialogue systems. However, reward models are often treated as task-specific modules, with limited examination of their validity or consistency with psychotherapy principles. In particular, little work systematically evaluates whether reward models capture long-horizon therapeutic goals. This gap motivates the need for dedicated benchmarking of reward models in long-horizon CBT-based psychological counseling settings.

\subsection{Reward Model}
Reward models (RMs) are a core component of large language model (LLM) alignment, providing preference signals that guide models toward human-preferred responses \cite{10.5555/3600270.3602281, Askell2021AGL, Bai2022TrainingAH}. They can be broadly divided into discriminative and generative RMs \cite{Zhang2024GenerativeVR, Zhou2024RMBCB}. Discriminative RMs are typically trained on human preference datasets to output scalar scores reflecting relative response quality, and are commonly used as optimization targets in reinforcement learning from human feedback (RLHF) pipelines \cite{10.5555/3600270.3602281, lambert-etal-2025-rewardbench, 10.5555/3294996.3295184}. Generative RMs, often referred to as LLM-as-a-judge \cite{Zhu2023JudgeLMFL, NEURIPS2023_91f18a12, kim-etal-2024-prometheus}, leverage powerful pretrained LLMs to directly evaluate responses via prompting. By providing explicit evaluation criteria, these models can produce preference judgments or detailed rationales without additional fine-tuning. Recent work has demonstrated that strong closed-source models can rival traditional discriminative RMs on general-domain preference tasks \cite{10.5555/3692070.3694459, Sun2023SALMONSW}. However, their reliability in domain-specific and long-horizon evaluation settings remains less understood.

\subsection{Benchmarks of Reward Models}
Several benchmarks have been introduced to systematically evaluate reward model reliability. Early efforts, such as RewardBench \cite{lambert-etal-2025-rewardbench}, provided a unified infrastructure for testing reward models across diverse domains, from open-ended chat to reasoning, and helped establish reward modeling as a research field. RMB \cite{Zhou2024RMBCB} propose the Best-of-N (BoN) evaluation as a new benchmark paradigm for assessing RMs. Subsequent efforts have expanded evaluation to more challenging or specialized settings, such as multilinguality \cite{gureja-etal-2025-rewardbench}, mathematical reasoning \cite{Kim2024EvaluatingRO}, and agentic behavior \cite{L2025AgentRewardBenchEA}. RoleRMBench \cite{Ding2025RoleRMBenchRoleRMTR} introduce the first benchmark for reward modeling in role-playing dialogue, where human preferences are nuanced, multi-faceted, and context-dependent. Despite their comprehensiveness, existing benchmarks predominantly focus on short-context or brief multi-turn interactions, leaving long-horizon preference modeling largely underexplored.

\begin{figure*}[!t]
\centerline{\includegraphics[width=\linewidth]{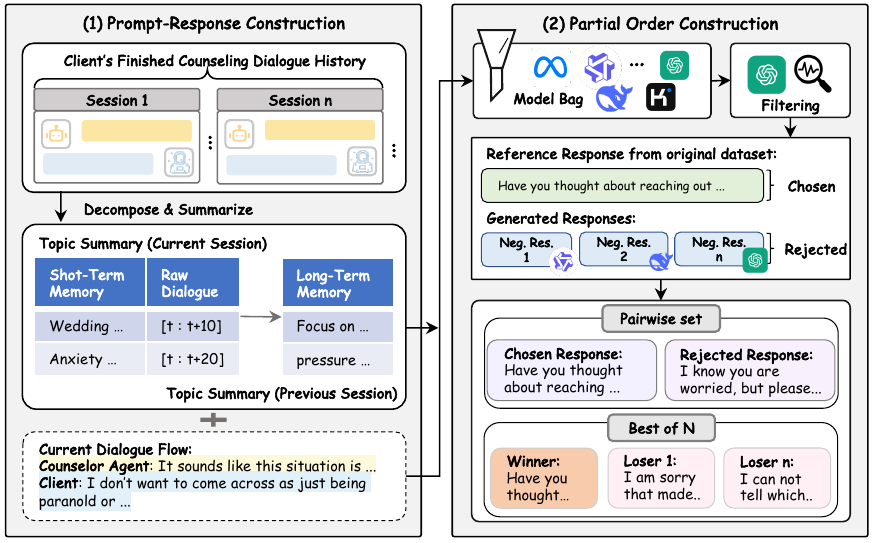}}
\caption{An overview of data construction process: (1) Sampling multi-session dialogues into prompts and obtaining multiple responses for them. (2) Organizing them into pairs or best-of-N lists.}
\label{fig:framework}
\end{figure*}
\section{Data Construction}
\label{sec:data_construction}
Our goal is to construct a benchmark that evaluates whether reward models (RMs) can serve as reliable proxies for CBT counselor preferences in long-horizon counseling scenarios, and whether they can provide effective reward signals for downstream tasks. Following prior work, we construct two types of partial ordering: a pairwise preference set with (\textit{chosen}, \textit{rejected}) response pairs, and a Best-of-N (BoN) set with (\textit{query}, \textit{winner}, \textit{losers}) tuples, where an RM selects the best response among multiple candidates. Fig.~\ref{fig:framework} provide an overall of the construction pipeline.

\subsection{Prompt-Response Construction}
\subsubsection{Counseling Case Sourcing}
We collect CBT counseling cases from publicly available resources, including real-world examples from the American Psychological Association (APA) website\footnote{\url{https://www.apa.org/pubs/databases/psyctherapy/}}, CBT textbooks, and simulated multi-session cases from DiaCBT \cite{Zhou2025DiaCBTAL}, a large-scale synthetic dataset created under CBT-guided constraints. All cases are used exclusively for research and evaluation. Each case is screened to ensure coverage to core CBT stages, including problem formulation, identification of automatic thoughts, cognitive restructuring, and behavioral interventions. After filtering and verification, we retain 118 validated cases, each comprising six sessions.

\subsubsection{Prompt Design}
To simulate long-horizon CBT counseling, each prompt follows a unified format consisting of: (1) a task instruction defining the counselor role, (2) a long-term summary capturing cross-session client states, (3) a short-term summary of the current session so far, and (4) the most recent dialogue turns. We adopt a progressive summarization strategy \cite{lv-etal-2024-coggpt}, which incrementally integrates prior session information into the current context. This approach preserves key therapeutic trajectories (e.g., core beliefs and prior strategies) while maintaining manageable context length. The prompt used are provided in the Appendix \ref{app:summary_generation}. In total, we construct over $13$k prompts spanning diverse long-horizon conditions.

\subsubsection{Response Generation} 
For each prompt, we generate counselor responses using ten state-of-the-art LLMs selected for their strong performance on mainstream leaderboards\footnote{\url{https://opencompass.org.cn/home}} and widespread use as baselines in recent studies. The models span diverse architectures and training paradigms. The complete list and generation settings are provided in the Appendix \ref{appendix:models_usage}. To create a robust preference benchmark, we generate two types of responses for each prompt: \textbf{Positive responses} are directly adopted from the original counselor utterances in the sourced CBT cases. These serve as clinically grounded chosen responses, reflecting established CBT practice, which act as stable, evidence-based anchors for preference signals, acknowledging that real therapeutic responses may vary but provide reliable reference points rooted in professional standards \cite{Chiu2024ACF}. \textbf{Negative responses} are generated via targeted prompt engineering on the same LLMs. We draw upon the qualitative meta-analysis by \cite{vybiral2024negative}, which identifies 21 client-reported negative experience meta-categories across four clusters (therapists' misbehavior, hindering relationship aspects, poor treatment fit, and negative treatment impacts). Negative variants start from a positive baseline prompt but introduce controlled, subtle deviations to embody one or more categories. Deviations are designed to be hidden (apparent only upon expert scrutiny) to simulate realistic, cumulative micro-harms in psychotherapy without overt toxicity.

\subsubsection{Response Filtering}
We apply a multi-stage filtering process to retain only high-quality, comparable responses. Candidates are removed if they are incomplete, off-topic, excessively repetitive, or contain unsafe content. We further exclude responses with extreme length deviations relative to reference responses from real CBT cases using rule-based checks. This filtering mitigates verbosity-related bias, ensuring that preference judgments reflect counseling quality rather than superficial length differences.

\subsection{Partial Order Construction}
\subsubsection{Pairwise Preference Construction} 
For each prompt, we form preference pairs by treating the original counselor response from the sourced case as the \textit{chosen} response and one model-generated negative response as the \textit{rejected} response. To increase difficulty and discriminative value, we retain only pairs with moderate quality gaps, discarding trivially poor responses. Rejected responses are sampled across multiple LLMs to balance model distributions.

\subsubsection{Agreement with Human Preference} 
To validate the quality of our constructed preference pairs in the complex long-horizon CBT domain, we randomly sampled 200 pairs and invited three independent human annotators (licensed CBT practitioners) to independently select the preferred counseling response (ties were not allowed). The annotators showed strong alignment with our constructed pairs, selecting the chosen response in 86\%, 92\%, and 94\% of the pairs, respectively. Moreover, all three annotators reached full consensus in 84\% of the pairs, and the majority vote agreed with our expert labels in 96\% of cases. These results demonstrate that clear and consistent preference signals exist in the majority of instances, confirming the high reliability of our benchmark.

\subsubsection{Best-of-N Set Construction} 
We additionally create a BoN evaluation set using a Best-of-4 configuration. For each query, the reference response is designated as the \textit{winner}, with four model-generated responses (distinct from those used in the pairwise set) as \textit{losers}. This design tests whether reward models can reliably recover clinically grounded responses in competitive settings, rather than exhaustively capturing all possible high-quality strategies.

\subsection{Detailed Statistics of our Benchmark}
The benchmark comprises over 13000 prompts derived from 118 CBT cases (each with 6 sessions), generating a large set of pairwise preference pairs and Best-of-N (BoN) queries. Below we present the distribution across the 6 therapeutic sessions, the 14 core CBT strategies, and the 21 negative experience categories, highlighting the scale and balance of the dataset.
\begin{figure}
\centering
\includegraphics[width=\linewidth]{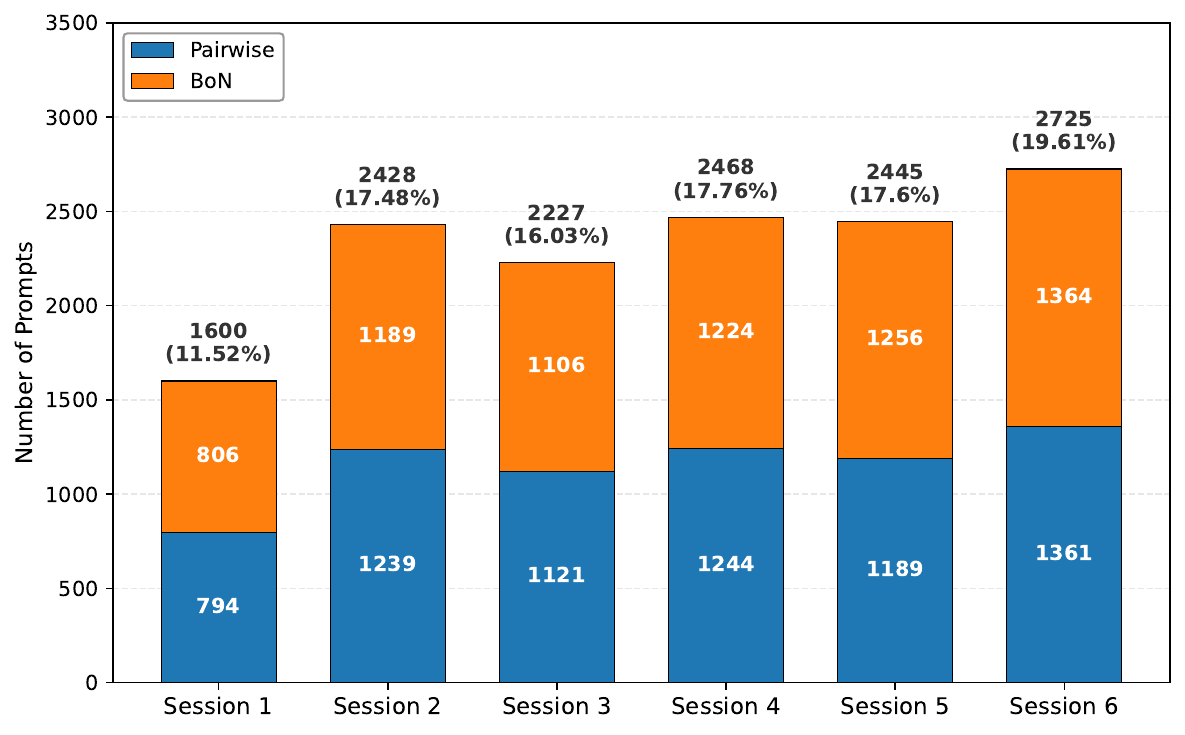}
\caption{Distribution of prompts (with percentages annotated above each bar), pairwise preference pairs, and Best-of-N queries across therapeutic sessions. Pairwise and BoN counts are shown inside the respective stacked segments. Session themes: (1) Information gathering, goal setting, and case conceptualization; (2) Identifying and challenging automated thoughts; (3) Discussion case conceptualization; (4) Adjusting intermediate beliefs and core beliefs; (5) Relapse prevention; (6) Consolidation.}
\label{fig:session_distribution}
\end{figure}
\subsubsection{Distribution across Sessions}
The benchmark contains 13893 prompts distributed across six sessions per case. As shown in Fig.~\ref{fig:session_distribution}, the number of prompts gradually increases from Session One to Session Six, with Session Six having the highest proportion (19.61\%). This distribution reflects the natural accumulation of therapeutic content and complexity in longer CBT interventions. The number of pairwise preference pairs (6948) and BoN queries (6945) maintains a relatively balanced allocation across sessions, ensuring robust evaluation across different therapeutic stages.

\begin{table}[!t]
\centering
\caption{Distribution of prompts, pairwise pairs, and Best-of-N queries across therapeutic strategies.}
\resizebox{\columnwidth}{!}{
\begin{tabular}{llcc}
\toprule
\textbf{Strategy} & \textbf{Prompts (\%)} & \textbf{Pairwise} & \textbf{BoN} \\
\midrule
Information Gathering & 811 (5.84) & 404 & 407 \\
Setting the Agenda & 580 (4.17) & 278 & 302 \\
Weekly Review & 411 (2.96) & 231 & 180 \\
Defining Therapeutic Objectives & 165 (1.19) & 85 & 80 \\
Psychoeducation & 1156 (8.32) & 571 & 585 \\
Working with Automatic Thoughts & 3973 (28.60) & 1990 & 1983 \\
Motivational Enhancement & 1905 (13.71) & 968 & 937 \\
Working with Intermediate and Core Beliefs & 799 (5.75) & 400 & 399 \\
Behavioral Techniques & 1087 (7.82) & 547 & 540 \\
Relapse Prevention & 553 (3.98) & 257 & 296 \\
Homework Assignments & 823 (5.92) & 419 & 404 \\
Requesting Feedback & 457 (3.29) & 227 & 230 \\
Summarization & 481 (3.46) & 239 & 242 \\
Other & 692 (4.98) & 332 & 360 \\
\midrule
\textbf{Total} & 13893 (100.00) & 6948 & 6945 \\
\bottomrule
\end{tabular}
}
\label{tab:strategy-distribution}
\end{table}
%
%
\subsubsection{Distribution across Core CBT Strategies}
Table~\ref{tab:strategy-distribution} summarizes the coverage of 14 core CBT strategies across the benchmark. The most frequently occurring strategy was ``\textit{Working with Automatic Thoughts}'' (28.60\%), followed by ``\textit{Motivational Enhancement}'' (13.71\%) and ``\textit{Psychoeducation}'' (8.32\%), which aligns with the core cognitive and preparatory elements of CBT. Less frequently occurring strategies, such as ``\textit{Defining Therapeutic Objectives}'' (1.19\%) and ``\textit{Relapse Prevention}'' (3.98\%), tended to appear in early or late treatment phases, consistent with typical therapeutic progression. These distributions ensure that the benchmark simultaneously captures long-horizon temporal dynamics and content diversity across specific CBT techniques, thereby providing a clinically representative testbed for assessing reward models' ability to detect subtle preference signals in psychotherapy.

\begin{figure}
\centering
\includegraphics[width=\linewidth]{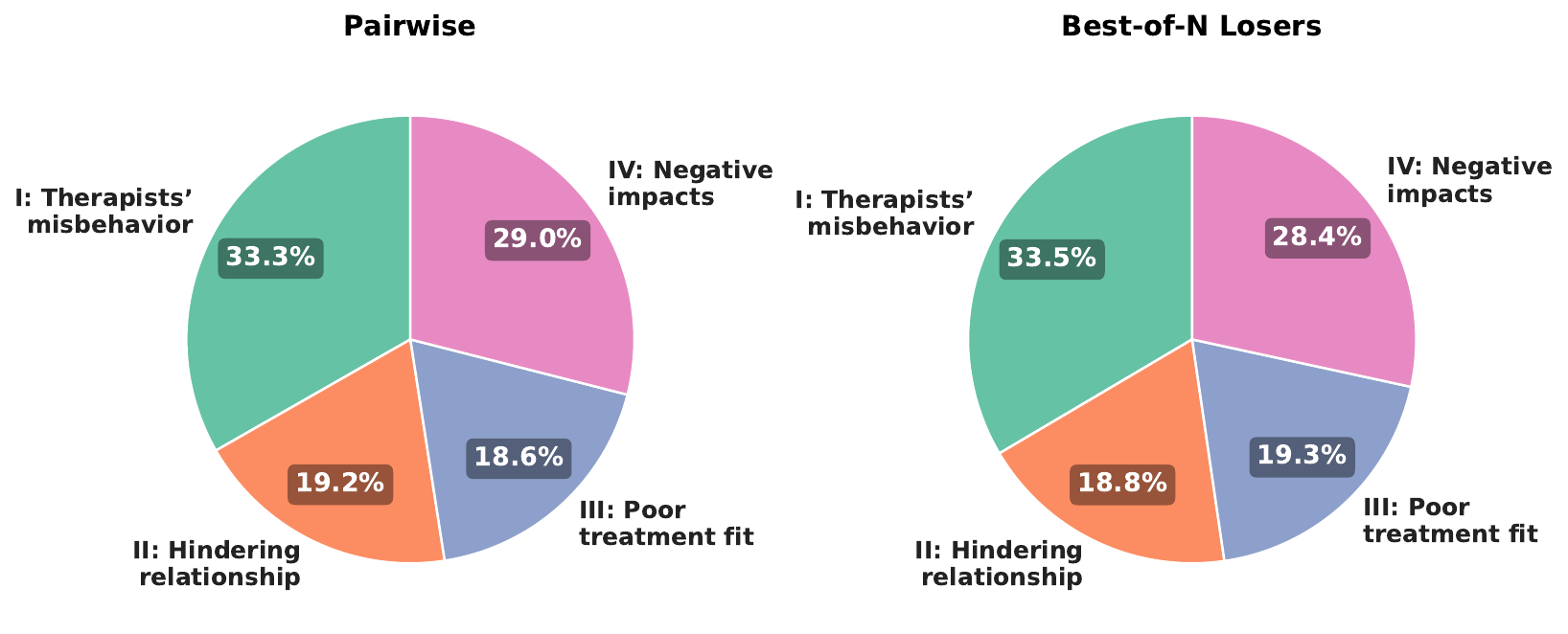}
\caption{Percentage distribution of negative experience categories in the pairwise and BoN loser sets, grouped by the four-cluster taxonomy.}
\label{fig:negative-clusters-pie}
\end{figure}
\begin{table}[!t]
\centering
\caption{Distribution of prominent sub-categories. Percentages are relative to the total negative experiences in each set (not within-cluster).}
\label{tab:cluster-sub-top}
\resizebox{\columnwidth}{!}{
\begin{tabular}{lcc}
\toprule
\textbf{Sub-category} & \textbf{Pairwise (\%)} & \textbf{BoN (\%)} \\
\midrule
\multicolumn{3}{c}{\textbf{Cluster \uppercase\expandafter{\romannumeral1}: therapists’ misbehavior}} \\
Therapist Not Listening & 4.63\% & 4.96\% \\
Therapist Not Understanding & 4.76\% & 4.67\% \\
Therapist Perceived as Incompetent & 4.59\% & 4.74\% \\
Therapist Devaluing the Client & 4.72\% & 4.86\% \\
Therapist Judging & 4.88\% & 4.76\% \\
Therapist Using Client for Own Benefit & 4.92\% & 4.73\% \\
Other Inappropriate Verbal Reactions & 4.78\% & 4.81\% \\
\midrule
\multicolumn{3}{c}{\textbf{Cluster \uppercase\expandafter{\romannumeral2}: hindering relationship}} \\
Distance and/or Lack of Empathy & 4.94\% & 4.61\% \\
Insecurity or Distrust & 4.56\% & 4.72\% \\
Confusion or Uncertainty & 5.04\% & 4.81\% \\
Poor Interpersonal Match with the Therapist & 4.63\% & 4.62\% \\
\midrule
\multicolumn{3}{c}{\textbf{Cluster \uppercase\expandafter{\romannumeral3}: poor treatment fit}} \\
Negative Evaluation of Practical Aspects & 4.43\% & 4.71\%  \\
Unmet Expectations & 4.56\% & 4.81\% \\
Lack of Fit with the Intervention & 4.88\% & 4.76\% \\
Dissatisfaction with Therapy Ending & 4.71\% & 4.99\% \\
\midrule
\multicolumn{3}{c}{\textbf{Cluster \uppercase\expandafter{\romannumeral4}: negative impacts of treatment}} \\
No Change or Insufficient Change & 4.88\% & 4.75\%  \\
Increased Problems after Therapy & 5.38\% & 4.68\% \\
Feeling Fear of the Therapy Process & 4.99\% & 4.73\% \\
Loss of Motivation or Hope & 4.53\% & 4.75\% \\
Unpleasant Feelings During Therapy & 4.69\% & 4.75\%\\
Negative Cognitions Aroused in Therapy & 4.48\% & 4.77\% \\
\bottomrule
\end{tabular}
}
\end{table}
\subsubsection{Distribution of Negative Experience Categories}
The distribution of negative experience categories is summarized at the cluster level in Fig.~\ref{fig:negative-clusters-pie}, following the four-cluster taxonomy proposed by \cite{vybiral2024negative}. In both the pairwise and BoN sets, \textit{Therapists’ misbehavior} (Cluster~I) is the most prevalent (33.3\% and 33.5\%, respectively), followed by \textit{Negative impacts of treatment} (Cluster~IV, 29.0\% and 28.4\%). Clusters~II and III (hindering relationship aspects and poor treatment fit) exhibit relatively lower but still substantial representation. A more detailed breakdown of prominent sub-categories within each cluster is provided in Table~\ref{tab:cluster-sub-top}. The most frequent sub-categories in the pairwise set are \textit{Increased Problems after Therapy} (5.38\%), \textit{Confusion or Uncertainty} (5.04\%), and \textit{Feeling Fear of the Therapy Process} (4.99\%). In the BoN set, the top three are \textit{Therapist Not Listening} (4.96\%), \textit{Therapist Devaluing the Client} (4.86\%), and \textit{Other Inappropriate Verbal Reactions} (4.81\%). This balanced yet clinically meaningful distribution across the four clusters, with relatively uniform sub-category contributions within each, ensures that the benchmark adequately captures the breadth of client-reported negative experiences, thereby enabling robust evaluation of reward models' sensitivity to subtle therapeutic harms.

\begin{table*}[!t]
\centering
\small
\caption{
Comparison of representative reward model benchmarks.
}
\begin{tabular}{lccccc}
\toprule
\textbf{Benchmark} &
\textbf{Best-of-N (N $>$ 2)} &
\textbf{Multi-session} &
\textbf{Unseen Prompts} &
\textbf{Target Skill} \\
\midrule
RewardBench \cite{lambert-etal-2025-rewardbench} 
& \xmark & \xmark & \xmark & General \\
RMB \cite{Zhou2024RMBCB} 
& \cmark & \xmark & \xmark & General \\
RewardBench2 \cite{Malik2025RewardBench2A} 
& \cmark & \xmark & \cmark & General \\
RoleRMBench \cite{Ding2025RoleRMBenchRoleRMTR} 
& \xmark & \xmark & \xmark & Role-playing \\
\midrule
\textbf{PRMB (ours)} 
& \cmark & \cmark & \cmark & CBT Counseling \\
\bottomrule
\end{tabular}
\label{tab:benchmark_comparison}
\end{table*}
\subsection{Comparison with Existing Benchmarks}
PRMB differs from existing RM benchmarks in both task structure and evaluation focus, as summarized in Table~\ref{tab:benchmark_comparison}. RewardBench, RMB, and RewardBench2 primarily evaluate preferences under short-context settings, targeting general-purpose skills such as factual correctness. Although effective for broad dialogue evaluation, these benchmarks do not capture the process-oriented nature of counseling interactions. Similarly, RoleRMBench focuses on role-playing consistency within a single interaction, but does not model session-level progression preference dependencies.

In contrast, PRMB is designed for long-horizon, multi-session CBT counseling. Preferences in PRMB depend on session-level coherence across counseling progression. Moreover, the progressive summarization framework distinguishes PRMB from prior benchmarks by preserving essential historical information while avoiding full dialogue transcripts, enabling realistic long-context evaluation at scale without exceeding context limits.

\section{Benchmarking Reward Models}
\label{sec:benchmarking}
We evaluate a broad range of state-of-the-art reward models (RMs) on PRMB. This section presents the evaluation setup and the main result.
\begin{table*}[!t]
\centering
\caption{The leaderboard of PRMB, ranked by the average of pairwise and BoN accuracy. $\star$ denotes LLM-as-a-judge models.}
\begin{tabular}{l|ccc}
\toprule
\textbf{Reward Model} & \textbf{Pairwise Acc.} & \textbf{BoN Acc.} & \textbf{Overall Acc.} \\
\hline
Llama-3.1-8B-Instruct-RM-RB2 &	0.8646 & 0.6700 & 0.7673 \\
deepseek-v3.2-exp$\star$ & 0.7796 & 0.7344 & 0.7570 \\
gpt-4o-mini$\star$ & 0.8368 & 0.6266 & 0.7317 \\
Skywork-Reward-V2-Qwen3-8B & 0.8365 & 0.6092 & 0.7229 \\
Skywork-Reward-V2-Qwen3-4B & 0.8247 & 0.5885 & 0.7066 \\
internlm2-20b-reward & 0.8122 & 0.5758 & 0.6940 \\
internlm2-7b-reward & 0.7974 & 0.5675 & 0.6825 \\
Skywork-Reward-V2-Qwen3-1.7B & 0.8051 & 0.5454 & 0.6753 \\
Qwen2.5-7B-Instruct$\star$ & 0.7860 & 0.5584 & 0.6722 \\
Gemma3-4B-IT$\star$ & 0.7533 & 0.5006 & 0.6270 \\
Skywork-Reward-V2-Qwen3-0.6B & 0.7596 &	0.4825 & 0.6211 \\
gpt-oss-20b$\star$ & 0.7500 & 0.4862 & 0.6181 \\
Mistral-7B-Instruct-v0.3$\star$ & 0.7306 & 0.4937 & 0.6122 \\
Skywork-Reward-V2-Llama-3.1-8B & 0.7306 & 0.4556 & 0.5931 \\
internlm2-1.8b-reward & 0.7047 & 0.3950 & 0.5499 \\
LLaMA-3.2-3B-Instruct$\star$ & 0.6513 & 0.4001 & 0.5257 \\
\bottomrule
\end{tabular}
\label{tab:leaderboard}
\end{table*}

\subsection{Evaluation Setup}
We evaluate both discriminative RMs and generative models under the LLM-as-a-Judge paradigm, assessing their ability to capture counselor-aligned preferences in long-horizon CBT counseling. Discriminative RMs assign a scalar reward to each prompt–response pair, following their standard role in RLHF pipelines. Generative RMs are prompted to directly select the preferred response among candidates following prior LLM-as-a-judge evaluations \cite{Zhou2024RMBCB}.

For each pairwise instance $i$, the benchmark provides a chosen and rejected response $(x_i^{+},x_i^{-})$. A discriminative RM is correct when it assigns a higher score to the chosen response. A generative RM is considered correct when it explicitly selects the chosen response. The pairwise accuracy $\mathcal{A}_{\text{pw}}$ is computed as:
\begin{equation}
    \mathcal{A}_{\text{pw}} = \frac{1}{N} \sum_{i=1}^{N}g(x_i^{+},x_i^{-}),
\end{equation}
where $g(\cdot)$ is $1$ if the RM prefers chosen over rejected response, otherwise is $0$.

For each BoN instance $i$, the benchmark provides a winner $x_i^{\star}$ and a set of losers $\{x_{ij}^{-}\}_{j=1}^{P_i}$. An RM succeeds on instance $i$ only if it prefers the winner over all losers. The BoN accuracy $\mathcal{A}_{\text{BoN}}$ is therefore:
\begin{equation}
    \mathcal{A}_{\text{BoN}} = \frac{1}{N} \sum_{i=1}^{N} \prod_{j=1}^{P_i} g(x_i^{\star},x_{ij}^{-}).
\end{equation}

We adopt these accuracy-based metrics to directly measure the correctness of relative preference judgments, which is the core requirement for RMs to recover grounded rankings and serve as reliable training signals.

\subsection{Evaluation Results}
\subsubsection{Comparison across Reward Models}
Table \ref{tab:leaderboard} reports the performance of all evaluated reward models, ranked by the average of pairwise and BoN accuracy. The leaderboard reveals a clear performance gradient on PRMB, with the strongest models achieving pairwise accuracies in the mid-to-high 80\% but BoN accuracies rarely exceeding 70\%, resulting in overall scores ranging from 52.57\% to 76.73\% , which substantially below typical ceilings on general-purpose reward modeling benchmarks.

The top-performing model, \textsc{Llama-3.1-8B-Instruct-RM-RB2}, attains 86.46\% pairwise accuracy and 67.00\% BoN accuracy (overall 76.73\%). The second-ranked model, \textsc{deepseek-v3.2-exp} (an LLM-as-a-judge model), follows closely with 77.96\% pairwise and 73.44\% BoN accuracy (overall 75.70\%). The \textsc{Skywork-Reward-V2} family demonstrates strong consistency and monotonic scaling: performance improves reliably with capacity (8B $>$ 4B $>$ 1.7B $>$ 0.6B), with pairwise accuracies ranging from 75.96\% to 83.65\% and BoN accuracies from 45.25\% to 60.92\%. Scaling trends in the \textsc{internlm2-reward} family are non-monotonic but overall positive: pairwise accuracy increases from 70.47\% (1.8B) to 79.74\% (7B) to 81.22\% (20B), and overall accuracy rises from 54.99\% to 68.25\% to 69.40\%, although gains diminish substantially from 7B to 20B. This pattern indicates that while larger capacity generally contributes to better alignment in this domain, returns to scale become marginal beyond mid-size models, potentially due to the specialized, process-sensitive nature of long-horizon CBT preferences.

Generative LLM-as-a-judge models exhibit mixed but generally limited transferability. High-capacity judges such as \textsc{gpt-4o-mini} (83.68\% pairwise, 62.66\% BoN) and \textsc{Qwen2.5-7B-Instruct} (78.60\% pairwise, 55.84\% BoN) perform respectably in pairwise comparisons but drop markedly in BoN, often approaching or falling below 60\%. This suggests that generic evaluative reasoning does not readily generalize to the nuanced, consistency-oriented preference signals required in long-term psychotherapy.

Overall, PRMB exposes challenges for both discriminative and generative reward models, highlighting gaps in their ability to maintain therapeutic preference alignment over extended interactions and against multiple subtle negative distractors.

\subsubsection{Comparison across Pairwise and BoN}
A substantial and consistent performance gap exists between the pairwise and BoN evaluation protocols across nearly all models.

Pairwise accuracy ranges from 65.13\% (\textsc{LLaMA-3.2-3B-Instruct}) to 86.46\% (\textsc{Llama-3.1-8B-Instruct-RM-RB2}), with most models clustered between 73\% and 84\%. In contrast, BoN accuracy spans a lower range from 39.50\% (\textsc{internlm2-1.8b-reward}) to 73.44\% (\textsc{deepseek-v3.2-exp}), with the majority falling between 45\% and 67\%. The average pairwise–BoN gap across all models exceeds 20 percentage points, and in many cases approaches 30 points.

This pronounced degradation in the BoN setting indicates that pairwise evaluation may overestimate RM alignment capability in realistic therapeutic contexts. While models can often reliably distinguish a positive reference response from a single negative variant in direct comparison, they frequently fail to rank the reference as the best when presented with four competing negative responses, many of which embody subtle harms drawn from the psychotherapy practice. The gap is particularly large for generative LLM-as-a-judge models, whose BoN scores frequently hover near or below 60\%, suggesting limited ability to maintain preference consistency amid multiple distractors.

\begin{figure*}[!t]
\centering
\includegraphics[width=\linewidth]{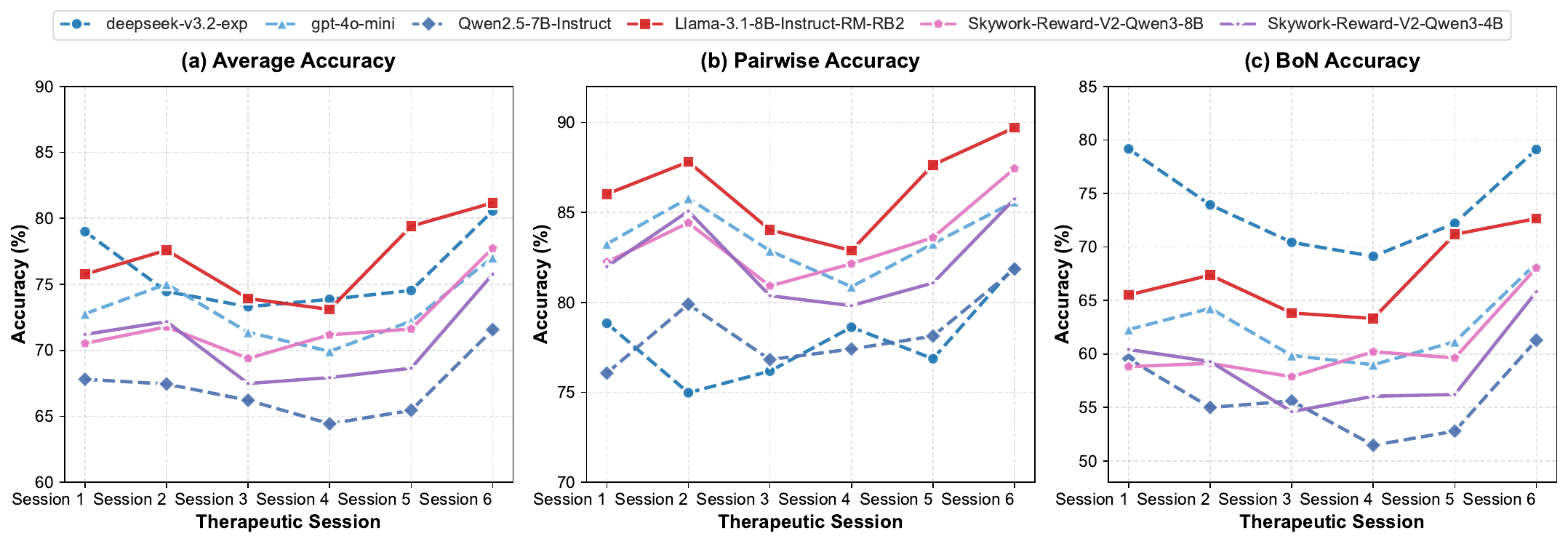}
\caption{(a) Average accuracy (pairwise + BoN), (b) Pairwise accuracy, and (c) Best-of-N accuracy across the six therapeutic sessions. Generative LLM-as-a-judge models (dashed lines) and discriminative reward models (solid lines) are distinguished by color and marker style. Solid lines represent discriminative reward models; dashed lines represent generative LLM-as-a-judge models.}
\label{fig:session-accuracy}
\end{figure*}
\subsection{Performance across Sessions}
To examine how reward models perform across different stages of CBT treatment, we report pairwise accuracy, BoN accuracy, and their average for six representative models on prompts stratified by the six therapeutic sessions. The selected models include the top three generative LLM-as-a-judge models (\textsc{deepseek-v3.2-exp}, \textsc{gpt-4o-mini}, \textsc{Qwen2.5-7B-Instruct}) and the top three discriminative reward models (\textsc{Llama-3.1-8B-Instruct-RM-RB2}, \textsc{Skywork-Reward-V2-Qwen3-8B}, \textsc{Skywork-Reward-V2-Qwen3-4B}).

Fig.~\ref{fig:session-accuracy} illustrates the average accuracy (pairwise + BoN) of these models across the six sessions. Performance varies by therapeutic stage, with early sessions (1–2) generally yielding higher BoN accuracies, likely due to simpler assessment, rapport-building, and agenda-setting content. Mid-to-late sessions (3–5) exhibit greater degradation in BoN performance for most models, reflecting the increasing complexity of cognitive restructuring, behavioral experiments, homework review, and consolidation phases.

Discriminative reward models (solid lines) consistently outperform generative LLM-as-a-judge models (dashed lines) across all sessions. \textsc{Llama-3.1-8B-Instruct-RM-RB2} achieves the highest overall performance, with average accuracy improving from early sessions (~75.8\%) to late sessions (~81.2\%), suggesting better handling of accumulated therapeutic context and process-oriented preferences in later stages. \textsc{Skywork-Reward-V2} variants maintain relatively stable performance, with modest improvement in later sessions.

In contrast, generative judges show larger pairwise–BoN gaps (average 15–20 percentage points) and more pronounced drops in mid-to-late sessions. For example, \textsc{gpt-4o-mini}’s BoN accuracy falls from 62.3\% (Session 1) to 58.9\%–68.5\% in Sessions 4–6, indicating limited robustness when multiple negative variants are present as treatment progresses.

These stage-dependent patterns highlight the increasing difficulty of therapeutic decision-making as CBT moves from initial problem formulation to behavioral consolidation and relapse prevention. The results suggest that current reward models remain more effective at coarse preference discrimination than at robust recovery of therapeutically optimal responses in long-horizon scenarios, particularly in later sessions where subtle, cumulative harms are more likely to emerge.

\subsection{Performance across CBT strategies}
To evaluate the effectiveness of different reward models in guiding LLM responses within CBT sessions, we compared the preference accuracy across 14 core CBT strategies. The selected models included the top-3 performing generative reward models and the top-3 discriminative reward models, as determined in prior evaluation stages.

\begin{figure}[!t]
\centering
\includegraphics[width=\columnwidth]{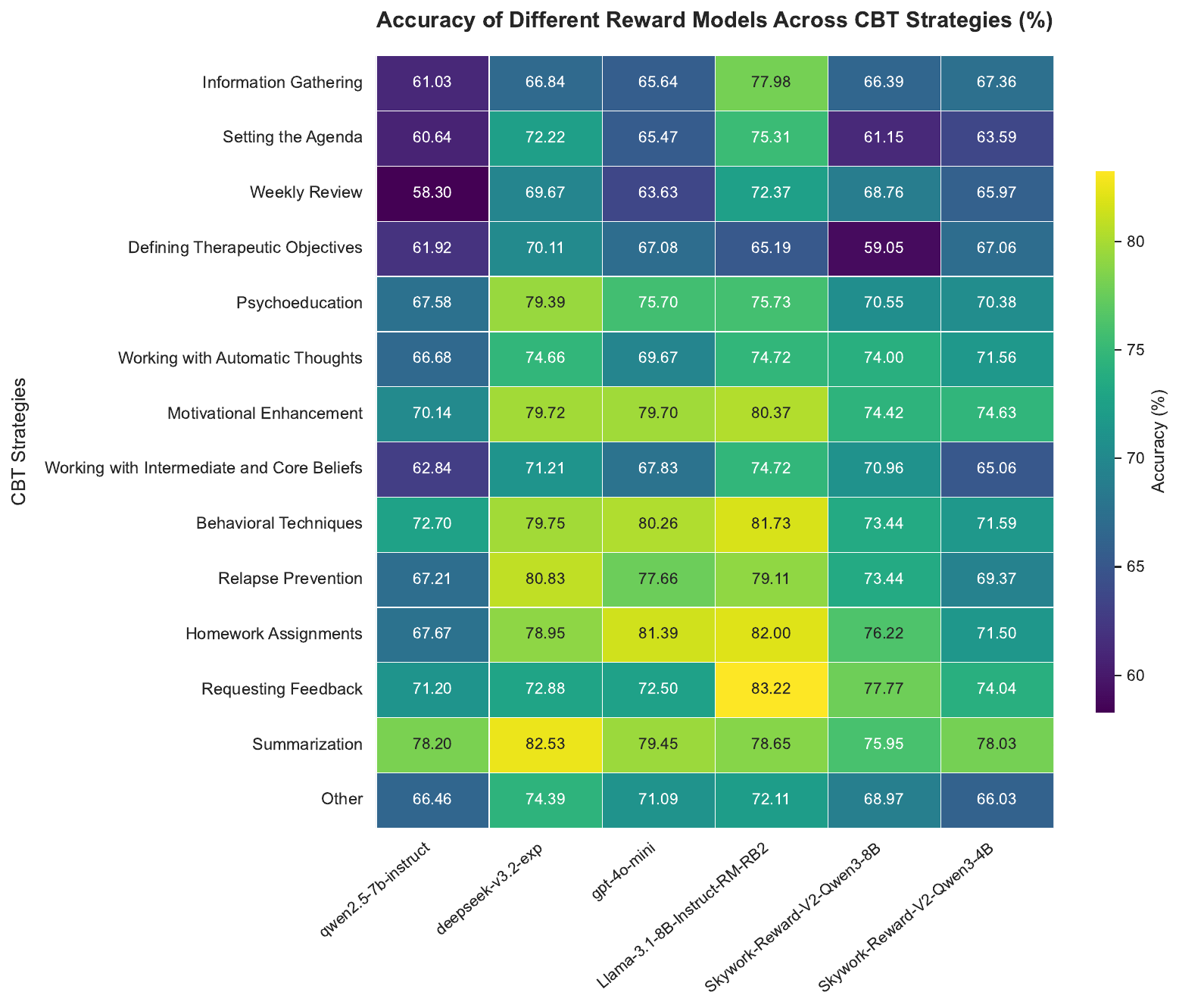}
\caption{Accuracy (\%) of top generative and discriminative reward models across CBT strategies. Values represent the percentage of average correct judgments for each strategy.}
\label{fig:cbt-strategy-accuracy}
\end{figure}

Fig.~\ref{fig:cbt-strategy-accuracy} presents the average accuracy (\%) of each model in correctly selecting the preferred response for instances labeled with each CBT strategy. Overall, the discriminative models achieved a marginally higher grand mean accuracy (72.39\%) compared to the generative models (71.50\%), although substantial variation existed both within and across model types.

The highest-performing model overall was \textsc{Llama-3.1-8B-Instruct-RM-RB2} (mean $=$ 76.66\%), which achieved the highest accuracy in eight out of the fourteen strategies, particularly excelling in homework assignments (82.01\%), requesting feedback (83.22\%), and behavioral techniques (81.74\%). The \textsc{deepseek-v3.2-exp} model ranked second (mean $=$ 75.23\%) and demonstrated particular strength in summarization (82.54\%), relapse prevention (80.83\%), and psychoeducation (79.39\%).

Notably, performance varied markedly across CBT strategies. Summarization, behavioral techniques, motivational enhancement, and homework assignments were among the highest-performing categories (grand means $>$ 76\%), suggesting that reward models more reliably distinguish high-quality responses in these structured, action-oriented domains. In contrast, strategies such as defining therapeutic objectives, setting the agenda, and weekly review showed comparatively lower and more variable performance (grand means $\approx$ 65$-$66\%), potentially reflecting greater subjectivity or context-dependence in these early-session or agenda-setting elements.

These findings indicate that discriminative reward models, particularly those fine-tuned with ranking-based objectives, may offer a modest but consistent advantage in CBT-specific preference modeling compared to generative reward approaches. However, the strong performance of certain generative models in specific high-level strategies highlights the potential value of generative reward models in future reward modeling for therapeutic dialogue systems.

\begin{figure}[!t]
\centering
\includegraphics[width=\columnwidth]{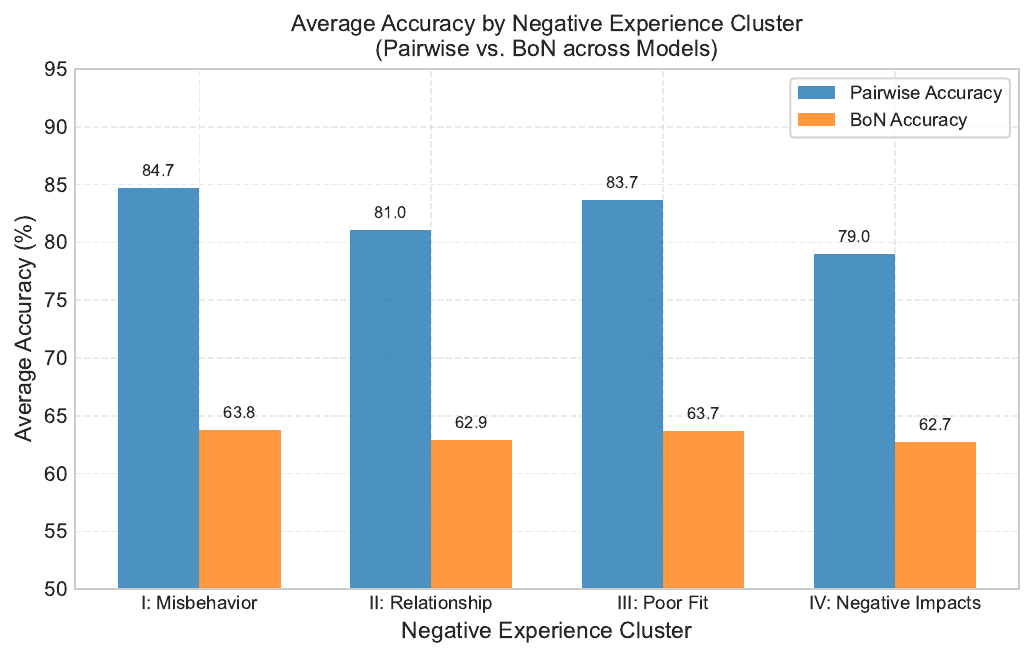}
\caption{Average pairwise (light bars) and BoN (dark bars) accuracy across the four main clusters of negative experiences. Values are averaged over the six evaluated reward models. Cluster I shows the highest performance, while Cluster IV exhibits the largest pairwise–BoN gap, highlighting challenges in detecting cumulative experiential harms.}
\label{fig:negative-cluster-performance}
\end{figure}
\subsection{Performance on Fine-grained Negative Experience}
To assess how reward models discriminate subtle negative experiences reported by psychotherapy clients, we aggregate the 21 meta-categories into their four main clusters and compute average pairwise and BoN accuracy across all instances within each cluster. 

Fig.~\ref{fig:negative-cluster-performance} shows the average pairwise and BoN accuracy across these four clusters for the top three generative and discriminative reward models. Detailed per-category pairwise and BoN accuracies for the six evaluated models are provided in Appendix~\ref{app:negative-performance}. Pairwise accuracy remains relatively high and stable across clusters (81–87\%), with the strongest performance on Cluster I (average ~85.8\%) and the lowest performance on Cluster IV(average 79.4\%). BoN accuracy is consistently lower (65–68\%), with the largest degradation observed in Cluster I (average ~20.9\%). Models show greater sensitivity to overt, immediate therapist errors than to subtle, process-oriented or cumulative harms. For instance, categories such as ``\textit{Therapist Judging}'', ``\textit{Therapist Devaluing the Client}'', and ``\textit{Therapist Not Listening}'' tend to yield higher BoN accuracies, while ``\textit{Increased Problems after Therapy}'', ``\textit{Loss of Motivation or Hope}'', and ``\textit{Unpleasant Feelings During Therapy}'' are among the most challenging, with BoN scores often falling below 65\% even for top-performing models.

These findings suggest that current reward models are more attuned to salient, observable misbehaviors than to experiential harms that accumulate over time and are central to client dissatisfaction in psychotherapy. The persistent BoN degradation across clusters underscores the need for reward modeling approaches that explicitly incorporate long-horizon reasoning and sensitivity to the full spectrum of negative therapeutic experiences.

\section{Downstream Evaluations}
\label{sec:downstream}
A core requirement of any RM benchmark is its ability to predict downstream alignment performance, rather than merely reflecting isolated preference judgments. In this section, we assess whether reward model rankings induced by PRMB correlate with real-world effectiveness under Best-of-N (BoN) inference, a widely used test-time strategy.

\subsection{Experimental Setup}
We assess downstream performance using a held-out set of counseling prompts. For each prompt, we sample 16 candidate responses from a policy model with temperature 0.8. Each reward model then scores and ranks the candidates, and the highest-scoring response is selected as the final output. To automatically evaluate generation quality, we compare the selected responses against a reference CBT counselor response using BERTScore \cite{bert-score}. Although BERTScore is a surface-level metric and does not fully capture therapeutic appropriateness, it provides a stable and reproducible proxy for relative quality differences, which is sufficient for evaluating ranking consistency across reward models. We then compute Spearman’s rank correlation coefficient between reward model rankings induced by PRMB and downstream rankings based on BERTScore. Correlations are computed by aggregating downstream BoN performance across four policy models\footnote{Qwen2.5-7B-Instruct, LLaMA-3.2-3B-Instruct, Mistral-7B-Instruct-v0.3, and Gemma3-4B-IT.}. This evaluation focuses on relative ranking consistency, rather than absolute generation quality.

\begin{figure}
\centering
\includegraphics[width=\linewidth]{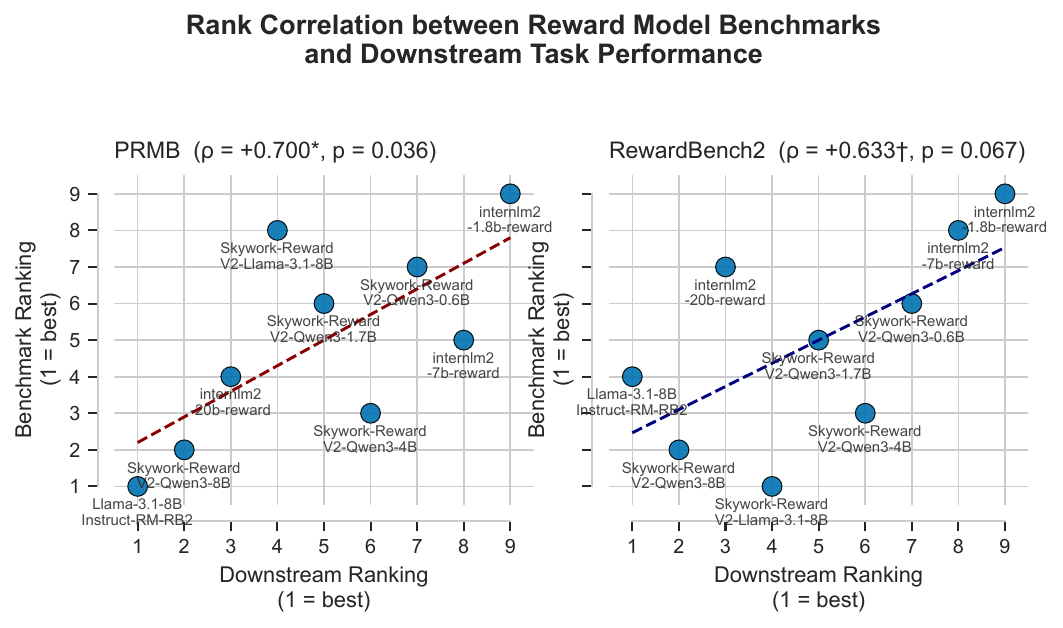}
\caption{Rank correlation between benchmark rankings (PRMB and RewardBench2) and downstream performance (ranked by average BERTScore of Best-of-N responses across four policy models; 1 = best). PRMB: $\rho = 0.700$ ($p \approx 0.036$); RewardBench2: $\rho = 0.633$ ($p \approx 0.067$).}
\label{fig:bon_corr}
\end{figure}
\subsection{Results}
Fig.~\ref{fig:bon_corr} shows the rank correlation between two automated reward model benchmark scores and downstream BoN inference sampling performance across four diverse policy models. The downstream ranking is derived from the average BERTScore of the BoN selected responses across four policy models, with 1 denoting the highest average BERTScore. PRMB shows a strong positive monotonic relationship with downstream performance (Spearman $\rho = 0.700$, $p \approx 0.036$), indicating that reward models ranked highly by PRMB consistently select responses exhibiting better lexical alignment (via BERTScore) to reference counselor responses. RewardBench2 exhibits a moderately strong correlation ($\rho = 0.633$, $p \approx 0.067$), which is directionally positive but only marginally significant. These findings suggest that PRMB offers more reliable predictive signal for inference-time alignment effectiveness in this long-horizon, therapeutic-domain task compared to the broader, general-purpose RewardBench2.

\section{Discussion}
\label{sec:discussion}
\begin{table}[!t]
\centering
\caption{Effects of inference-time strategies on pairwise and Best-of-N (BoN) accuracy across different generative reward model backbones.}
\begin{tabular}{lcc}
\toprule
\textbf{Method} & \textbf{Pairwise Acc.} & \textbf{BoN Acc.} \\
\midrule
gpt-4o-mini & 0.8368 & 0.6266 \\
+ Few shot (2-shot) & 0.8271 &  0.6000 \\
+ \textbf{RAG} & \textbf{0.8582} & \textbf{0.6584} \\
+ Self Refine & 0.8472 & 0.6288 \\
+ CoT & 0.8276 & 0.5783 \\
\midrule
Qwen2.5-7B-Instruct & 0.7860 & 0.5584 \\
+ Few shot (2-shot) & 0.7706 & 0.5379 \\
+ \textbf{RAG} & \textbf{0.8044} & \textbf{0.6029} \\
+ Self Refine & 0.7700 & 0.5479 \\
+ CoT & 0.7884 & 0.5054 \\
\midrule
LLaMA-3.2-3B-Instruct & 0.6513 & 0.4001 \\
+ Few shot (2-shot) & 0.6220 & 0.3865 \\
+ \textbf{RAG} & \textbf{0.6950} & \textbf{0.4813} \\
+ Self Refine & 0.4800 & 0.2071 \\
+ CoT & 0.5025 & 0.2535 \\
\bottomrule
\end{tabular}
\label{tab:discussion}
\end{table}
\subsection{Effects of Inference-Time Strategies}
Table~\ref{tab:discussion} presents the effects of several common inference-time strategies applied to generative reward models with different backbone architectures. The results reveal substantial variability in performance across model families and evaluation protocols (pairwise vs. Best-of-N), with no strategy delivering consistent improvements except Retrieval-Augmented Generation (RAG).

Few-shot prompting (2-shot) consistently degrades performance across all three backbones and both evaluation protocols. Pairwise accuracy decreases by 0.97\% on \textsc{gpt-4o-mini}, 1.54\% on \textsc{Qwen2.5-7B-Instruct}, and 2.93\% on \textsc{LLaMA-3.2-3B-Instruct}, with corresponding BoN declines of 2.66\%, 2.05\%, and 1.36\%. This uniform degradation suggests that providing a small number of in-context preference examples introduces noise or fails to align with the complex preference distributions in long-horizon therapeutic dialogues, regardless of model scale or architecture.

Self-Refine and Chain-of-Thought (CoT) prompting exhibit the most detrimental effects, particularly in the more challenging Best-of-N setting. On \textsc{gpt-4o-mini} and \textsc{Qwen2.5-7B-Instruct}, both methods cause small to moderate declines in pairwise accuracy and more pronounced drops in BoN accuracy. The degradation is especially severe for \textsc{LLaMA-3.2-3B-Instruct}: Self-Refine reduces pairwise accuracy to 48.00\% ($-$17.13\%) and BoN to 20.71\% ($-$19.30\%), while CoT reduces pairwise to 50.25\% ($-$14.88\%) and BoN to 25.35\% ($-$14.66\%). These sharp collapses indicate that iterative self-correction and verbose step-by-step reasoning can amplify initial errors or reinforce superficial heuristics rather than resolve subtle, trajectory-level preference misalignments typical of multi-turn therapeutic interactions.

Retrieval-Augmented Generation (RAG) stands out as the strategy that consistently improves performance across all three backbones and both evaluation protocols. It delivers the largest relative gains on \textsc{LLaMA-3.2-3B-Instruct} ($+$4.37\% pairwise, $+$8.12\% BoN), followed by \textsc{Qwen2.5-7B-Instruct} ($+$1.84\% pairwise, $+$4.45\% BoN) and \textsc{gpt-4o-mini} ($+$2.14\% pairwise, $+$3.18\% BoN). These consistent positive effects suggest that injecting external CBT-related knowledge effectively compensates for knowledge gaps and improves alignment with therapeutically optimal responses.

The reliable improvement from RAG highlights a promising direction: incorporating high-quality external reference feedback signals during inference may help generative reward models better capture process-oriented preferences in counseling dialogues. Unlike heuristic-based methods (few-shot, Self-Refine, CoT) that often introduce noise or amplify weaknesses, RAG provides domain-specific guidance that aligns more closely with the task's requirements. While the absolute gains remain modest and do not fully close the performance gap between open and closed models, these results motivate further exploration of reference-augmented approaches, such as preference-pair retrieval, session-context-aware RAG.

Collectively, the experiments indicate that popular post-training inference-time strategies offer no robust or reliable benefit for generative reward modeling in long-horizon, process-oriented counseling tasks. Among the tested methods, only RAG consistently yields positive (albeit limited) gains, while few-shot prompting, Self-Refine, and CoT frequently degrade performance, particularly in the Best-of-N setting and on resource-constrained models. These findings highlight the limitations of heuristic-based inference enhancements and underscore the need for training-time interventions, more sophisticated reference mechanisms, and evaluation benchmarks that better capture session-level preference dynamics and long-term therapeutic coherence.

\section{Conclusion}
\label{sec:conclusion}
In this paper, we propose PRMB, a benchmark designed to evaluate reward models in long-horizon, multi-session CBT-based counseling. Extensive evaluations show that both discriminative and generative reward models struggle to reliably capture counselor-aligned preferences in this setting. We additionally find that inference-time heuristics alone are insufficient for aligning reward models in process-oriented conversational domains. We hope PRMB will serve as a foundation for future research on reward modeling methods that explicitly account for long-horizon structure, relative judgment, and therapeutic trajectories.

\appendices
\section{Additional Materials on Data Construction}
\label{app:add_materials}

\subsubsection{Progressive Summary Generation}
\label{app:summary_generation}
\begin{figure}
    \centering
    \includegraphics[width=\linewidth]{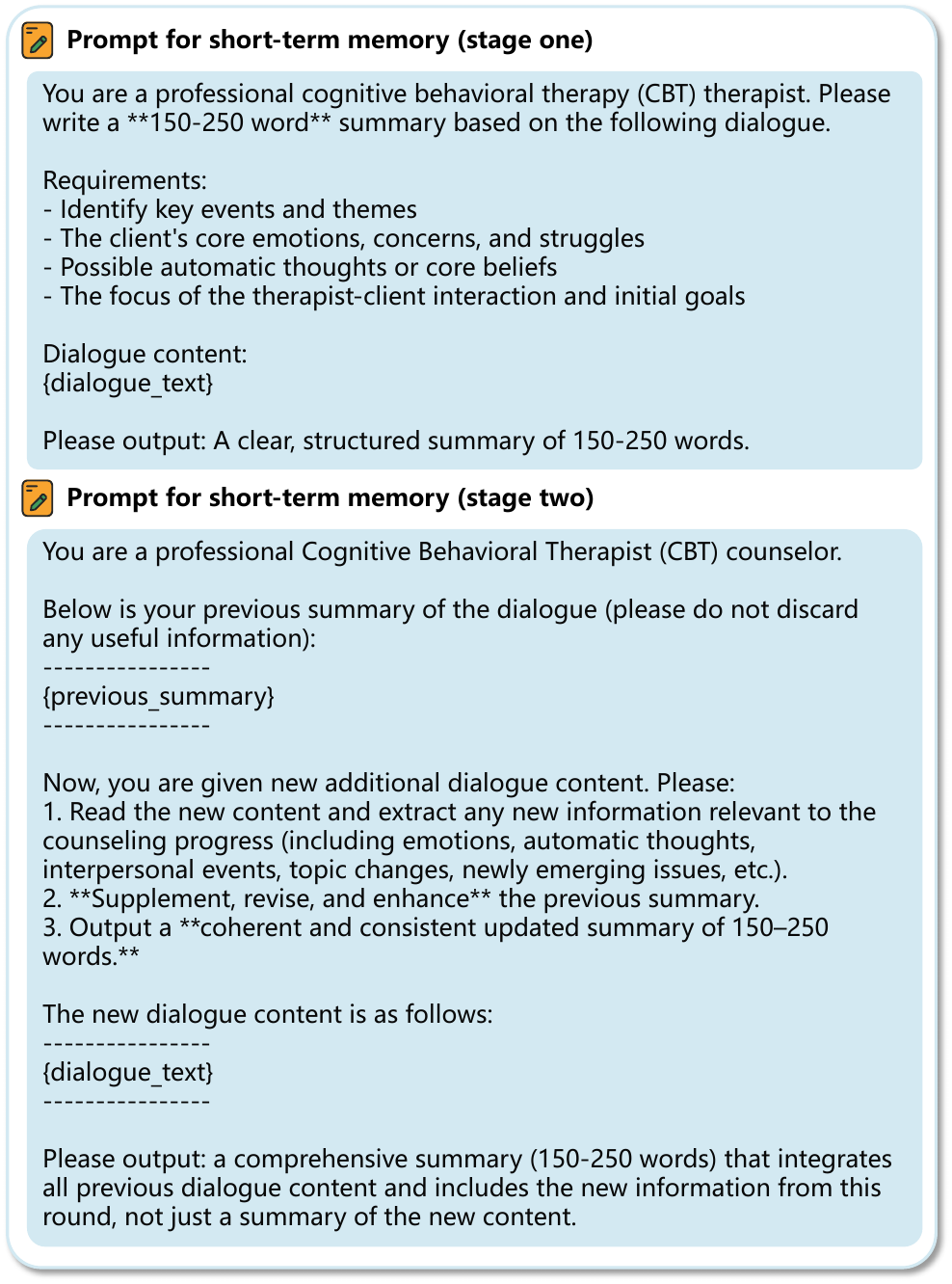}
    \caption{Prompt usage in short-term memory generation.}
    \label{fig:short_term_memory}
\end{figure}

\begin{figure}
    \centering
    \includegraphics[width=\linewidth]{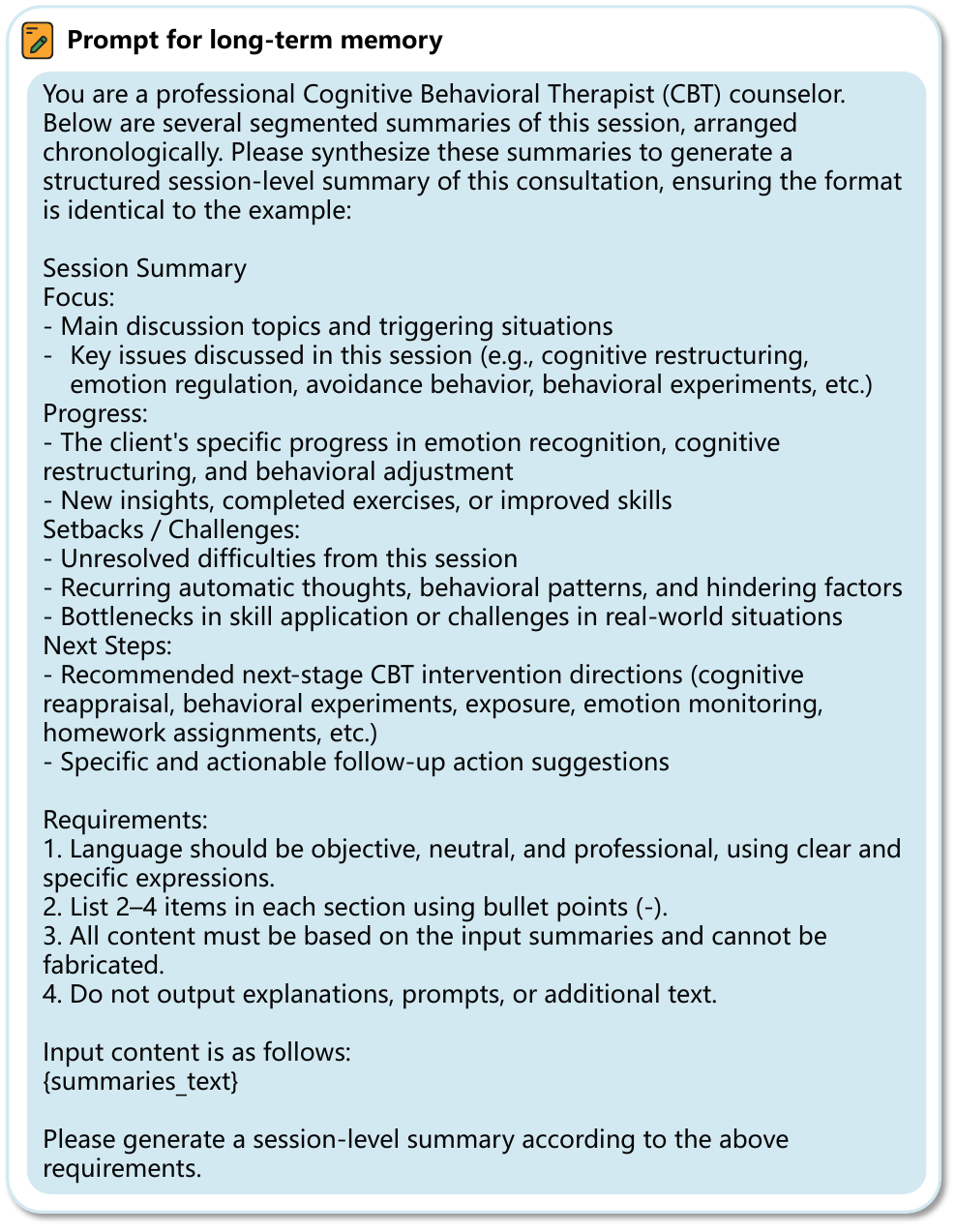}
    \caption{Prompt usage in long-term memory generation.}
    \label{fig:long_term_memory}
\end{figure}

To support long-horizon CBT counseling evaluation while keeping prompt length manageable, we adopt a progressive summary generation framework to construct counseling prompts. Historical information is incrementally compressed into structured summaries that are carried forward across sessions. Specifically, for each counseling case consisting of multiple sessions, we maintain two types of summaries:
\begin{itemize}
    \item \textbf{Short-term summary}, which captures the key developments within the current session up to the present turn, such as recent emotional states and newly expressed thoughts, as show in Fig. \ref{fig:short_term_memory}.
    \item \textbf{Long-term summary}, which captures stable and cross-session information about the client, including presenting problems, core beliefs, recurring automatic thoughts, previously identified cognitive distortions, and therapeutic strategies that have been introduced, as shown in Fig. \ref{fig:long_term_memory}.
\end{itemize}

At the beginning of the first session, both summaries are initialized as empty. After each session concludes, the short-term summary is merged into the long-term summary using a structured summarization prompt, and the short-term summary is reset for the next session. This process ensures that essential therapeutic trajectories are preserved across sessions without exceeding context length. Importantly, summaries are constructed solely from prior dialogue content and do not incorporate or reference any candidate counselor responses used for preference evaluation. This prevents information leakage and ensures that reward models are evaluated only on information that would be available in realistic long-horizon counseling settings.

\subsubsection{Model List Usage in Response Candidate Generation}
\label{appendix:models_usage}
The models used for response candidate generation include both open-weight and closed-source systems, and span instruction-tuned and chat-optimized variants. All models are used in a zero-shot generation setting with fixed decoding parameters to ensure comparability across candidates. Across all models, responses are generated with temperature set to $0.8$ and a maximum output length sufficient to complete a full counseling turn. No model-specific prompt tuning or post-processing is applied. Each prompt is paired with multiple independently generated responses sampled from different models, enabling the construction of challenging preference pairs and Best-of-N instances that reflect realistic model diversity. A complete list of models used for response candidate generation is shown in Table~\ref{tab:gen_models}.
\begin{table}[h]
\centering
\caption{Models used for counselor response candidate generation.}
\begin{tabular}{l l}
\toprule
\textbf{Model Name} & \textbf{Type} \\
\midrule
GLM-4.6 & Closed-source \\
Qwen3-235B-A22B-Thinking-2507 & Open-weight \\
Qwen3-30B-A3B-Thinking-2507 & Open-weight \\
Qwen3-32B & Open-weight \\
DeepSeek-V3.2-Exp & Open-weight \\
DeepSeek-R1-0528 & Open-weight \\
gpt-4o-mini & Closed-source \\
gpt-oss-20b & Open-weight \\
MiniMax-M2 & Open-weight \\
Kimi-K2 & Open-weight \\
\bottomrule
\end{tabular}
\label{tab:gen_models}
\end{table}

\section{Detailed Performance on Negative Experience}
\label{app:negative-performance}
\begin{figure}
\centering
\includegraphics[width=\linewidth]{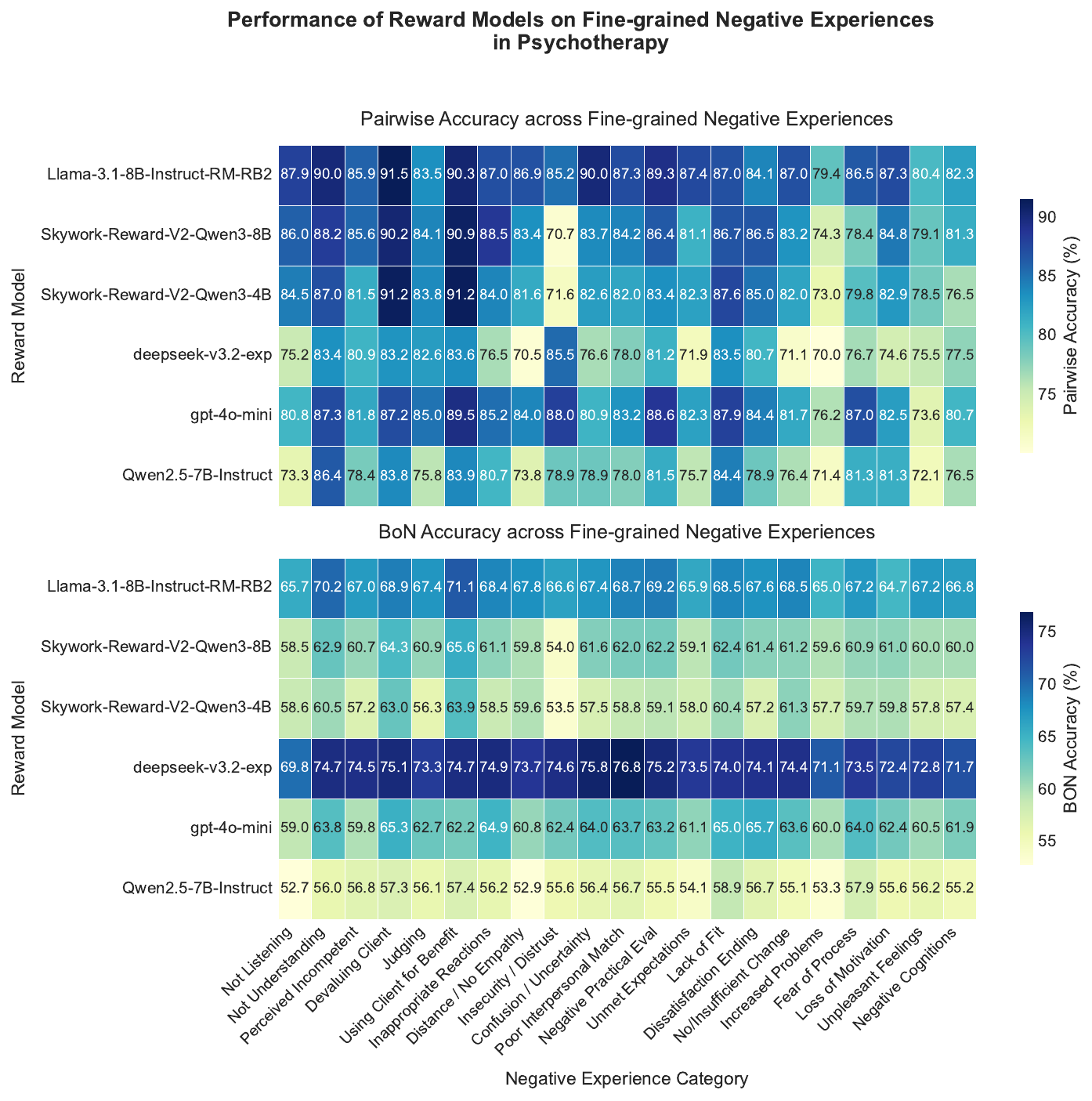}
\caption{Detailed per-category pairwise and BoN accuracies for the six evaluated models.}
\label{fig:negative-category-heatmap}
\end{figure}

This appendix provides detailed pairwise and BoN accuracy for each of the 21 negative experience meta-categories, across the six evaluated reward models. Categories are grouped by their original four-cluster taxonomy for reference. Fig.~\ref{fig:negative-category-heatmap} provides a heatmap visualization of BoN accuracy across models and categories, with darker shades indicating higher performance.

\bibliographystyle{IEEEtran}
\bibliography{TAC_Latex/my}


 





\end{document}